\documentclass{aastex62}

\received{January 20, 2020}
\revised{ }
\accepted{ }

\shorttitle{The Alpha Persei cluster and stream}
\shortauthors{Nikiforova et al.}

\begin{document}

\title{The relation of the Alpha Persei star cluster with the nearby stellar stream}

\correspondingauthor{Anton F. Seleznev}
\email{anton.seleznev@urfu.ru}

\author{Victoria V. Nikiforova}
\affil{Ural Federal University \\
620002, 19 Mira street, \\
Ekaterinburg, Russia}

\author{Maxim V. Kulesh}
\affil{Ural Federal University \\
620002, 19 Mira street, \\
Ekaterinburg, Russia}

\author{Anton F. Seleznev}
\affil{Ural Federal University \\
620002, 19 Mira street, \\
Ekaterinburg, Russia}

\author{Giovanni Carraro}
\affil{Dipartimento di Fisica e Astronomia, Universita'  di Padova\\
Vicolo Osservatorio 3\\
I35122, Padova, Italy}

\begin{abstract}
A map of 100 degrees on a side extracted from Gaia DR2 and centred on Alpha Persei reveals two
distinct structures --- the Alpha Persei star cluster and a conspicuous stellar stream, as widely documented in recent literature. In this work we employ
DBSCAN to assess individual stars' membership and
attempt at separating stars belonging to the cluster and to the stream from the general field. In turn, we characterize the stream and investigate its relation with the cluster.
The stream population turned out to be significantly older (5$\pm 1$Gyr) than the cluster, and to be positioned $\sim$90 pc away from the cluster, in its background. The stream exhibits a sizeable thickness of $\sim 180$ pc in the direction of the line of view. Finally,
the stream harbours a prominent population of white dwarf stars.
We estimated an upper limit of the stream mass of $\sim 6000 M_{\odot}$.
The stream would therefore be the leftover of a relatively massive old cluster.
The surface density map of the Alpha Persei evidences the presence of tidal tails.  While it is tempting to ascribe them to the interaction with the disrupting old star cluster, we prefer to believe, conservatively, they are of Galactic origin.
\end{abstract}


\section{Introduction}

The first mention of the Alpha Persei star cluster (Melotte 20, Collinder 39) is in  \citet{Edd}, who firstly noticed a group
of common motion bright stars in the vicinity of the star. \citet{Art} identified candidate kinematic members brighter than $B=12$ mag within
4.5 degrees from the cluster center. The analysis of the vector point diagram allowed her to isolate 163 stars within
$\sigma\sqrt2=10$ mas/yr from the mean proper motion, 287 stars within $2\sigma=14$ mas/yr, and 503 stars within $3\sigma=20$ mas/yr.
\citet{Art} investigated also the cluster projected density profile and its luminosity function. Of the 503 $3\sigma$ candidates she isolated
83 {\it bona fide} members. The cluster corona was computed as 7.3 pc
(2.5 degrees) and the core as 3.7 pc (1.25 degrees).

\citet{SRB} studied Alpha Persei super-corona  and extracted from the Smithsonian Astrophysical Observatory Star Catalog ($RA=2-4$ hours,  $DEC=+40-+60$ degrees)
stars with spectroscopic information (B stars down to V = 8, and A0-A3 stars down to V = 9 mag). The circle 2.5 degrees wide around the cluster center was therefore excluded. This super-corona was found to be composed of 93 B-A3 co-moving stars.

\citet{Makarov} investigated Alpha Persei with  Tycho-2
and the Second USNO CCD Astrographic Catalog (UCAC2). He found 139 candidate members and estimated an age of 52 Myr . \cite{Makarov}
also noticed that the cluster is surrounded by a sparse halo of co-moving dwarfs identified from 2MASS photometry.

The first suggestion that Alpha Persei cluster might be  associated with this extended group of common motion stars is in \citet{Mermilliod} who firstly introduced the term `stream'.

\citet{vanLeeuwen} selected 50 stars from the Hipparcos catalog 8 degrees around the cluster center and derived $\mu_{\alpha}$= 22.73$\pm$0.17 mas$/$yr, $\mu_{\delta}$= -26.51$\pm$0.17 mas$/$yr, $\pi$ =5.80$\pm$0.09 mas, (m-M)=  6.18$\pm$0.03 mag (hence 172.4$\pm$2.7 pc of distance), and log(age) = 7.55.
He concluded that the overall stars' properties characterize Alpha Persei as a remnant of an OB association more than as a bound cluster.

By following \cite{Perryman} and \citet{Luri},
\citet{Lodieu} investigated Alpha Persei with Gaia DR2 data. He inferred a tidal radius of 9.5 pc and found 554 stars within it. Extending the searching area to three tidal radii (28.5 pc) allowed him to find 2041 sources.
\citet{Lodieu} derived a distance of 177.68$\pm$0.84 pc, a tangential velocity of 28.7$\pm$0.5 km/s and a core radius of 2.3$\pm$0.3 pc containing 21 stars.

The Gaia DR2 data release \citep{2016gaia,2018gaia} allows one to adopt two different approaches to study star clusters.

The first approach consists of selecting stars with the most reliable parameters (parallaxes --- Plx, or $\pi$ ---,
proper motion components --- $\mu_{\alpha}$ and $\mu_{\delta}$ ---, and radial velocities).
Cluster members would then crowd in a 5-dimensional or 6-dimensional parameter space.
This approach permits a detailed investigation of
star clusters' 3-dimensional structure and internal kinematics \citep{jeffries,Vela,damiani,cantat}.
The limitation of this approach is that only cluster stars for which parameters are precisely measured can be used.

The second approach consists in a statistical investigation and makes use of the virtue of Gaia as an all-sky survey complete down to a specified limiting magnitude.
Within this statistical approach, the ultimate goal of the stars' selection is of increasing the contrast of the cluster against the field. We set the limits of the parameters (parallaxes and the proper motions) in such a way that probable cluster members are not missing, despite the possibly large errors of the parameters. At the same time, we meet the advantage that the number of field stars - and hence field stars density fluctuations - are kept low.

Then, statistical methods are used to evaluate the distribution functions
characterizing the cluster under study, such as surface or spatial densities, luminosity and mass function (see, for example, \citet{LF,
profiles,kernel}). The ultimate goal is to obtain complete realizations of the distribution functions in a statistical sense.

An illustration of the application of both approaches is in our recent study of the old nearby star cluster Ruprecht 147 \citep{Rup147}. The first approach revealed the tidal tails of the cluster while the second
one was used to derive  the radial density profile (with the cluster corona outside the cluster tidal radius), the luminosity and mass functions. The
first approach used 69 candidate members only given the very strict selection. From the application of the second approach the number of the cluster stars
were 280$\pm$67 and the cluster mass  234$\pm$52 $M_{\odot}$.

The goal of this paper is to study the area around Alpha Persei with the purpose of separating stars belonging to the cluster and to the stream from the general field
and for investigating their relationship.

As a consequence, this paper is organized as follows. Section 2 is devoted to a brief description
of Gaia DR2 sample.
In Section 3 we discuss the properties of the cluster and the stream and their mutual relationship. Section 4 is devoted to an analysis of the spatial distribution of white dwarf stars in the area.
Section 5, finally, summarizes our results.

\section{The sample}
We first selected stars from Gaia DR2 in the five-dimensional space defined as : $l\in[90;200]$ deg, $b\in[-55;45]$ deg, $\pi\in[3;8]$ mas, $\mu_{\alpha}\in[15;30]$ mas/yr, $\mu_{\delta}\in[-32;-18]$ mas/yr. This selection returned 60603 stars. We will refer to this selection as Sample 1a.

The text of the ADQL query was as follows:

{\tt SELECT l, b, ra, ra\_error, dec, dec\_error, source\_id, parallax, parallax\_error, pmra, pmra\_error, pmdec, pmdec\_error, phot\_g\_mean\_mag, phot\_bp\_mean\_mag, phot\_rp\_mean\_mag, bp\_rp, radial\_velocity, radial\_velocity\_error, phot\_bp\_mean\_flux, phot\_bp\_mean\_flux\_error,
phot\_rp\_mean\_flux,
phot\_rp\_mean\_flux\_error, phot\_g\_mean\_flux, phot\_g\_mean\_flux\_error
 {\raggedright

}
FROM gaiadr2.gaia\_source

WHERE b BETWEEN -55.0 and 45.0 AND l BETWEEN 90.0 and 200.0

AND pmra BETWEEN 15.0 and 30.0 AND pmdec BETWEEN -32.0 and -18.0

AND parallax BETWEEN 3.0 AND 8.0.}

The corresponding area on sky is about 100 degree on a side. Such wide area
would allow us to detect structures around the cluster possibly associated with the stream.  Since our goal is to study the cluster in a statistical manner,
in order to preserve the largest possible number of cluster members we did not use any quality flag. When we add the limitation on stellar magnitude $G<18$ mag, the sample diminishes to 32442 stars.
We will refer to this selection as Sample 1.
We adopted the fundamental parameters of Alpha Persei as in \citet{LP}, see Table 1.

\begin{table}
\normalsize
\bigskip
\begin{center}
\vspace{2 mm} Table 1. Fundamental parameters of Alpha Persei.

\vspace{2 mm}
\begin{tabular}{|l|c|}
\hline
Parameter              & Value             \\
\hline
Right Ascension        & $03^h27^m$        \\
Declination            & $+49^{\circ}07'$  \\
Galactic longitude     & $147.5^{\circ}$   \\
Galactic latitude      & $-06.5^{\circ}$   \\
Log (Age [yr])          & 7.9               \\
Heliocentric distance  & 176 pc            \\
Color excess E(B-V)    & 0.1 mag           \\
\hline
\end{tabular}
\end{center}
\end{table}

\begin{figure}
   \centering
   \includegraphics[width=8truecm]{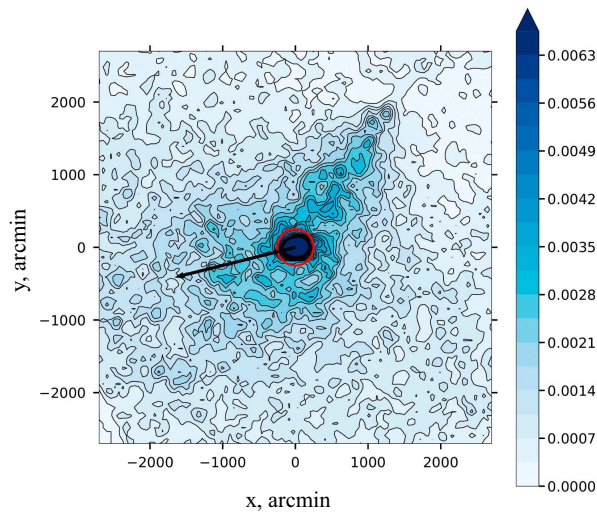}
   \caption{Surface density map for Sample 1. The red circle corresponds to Alpha Persei tidal radius, while the black arrow
   indicates the mean motion of the cluster stars (see below). The horizontal axis decreases with the Galactic longitude and  vertical axis rises with
   the Galactic latitude. The density values are in $\rm arcmin^{-2}$ (see their scale rightward of the map).}
   \label{map_S1}
   \end{figure}

\begin{figure}
\gridline{\fig{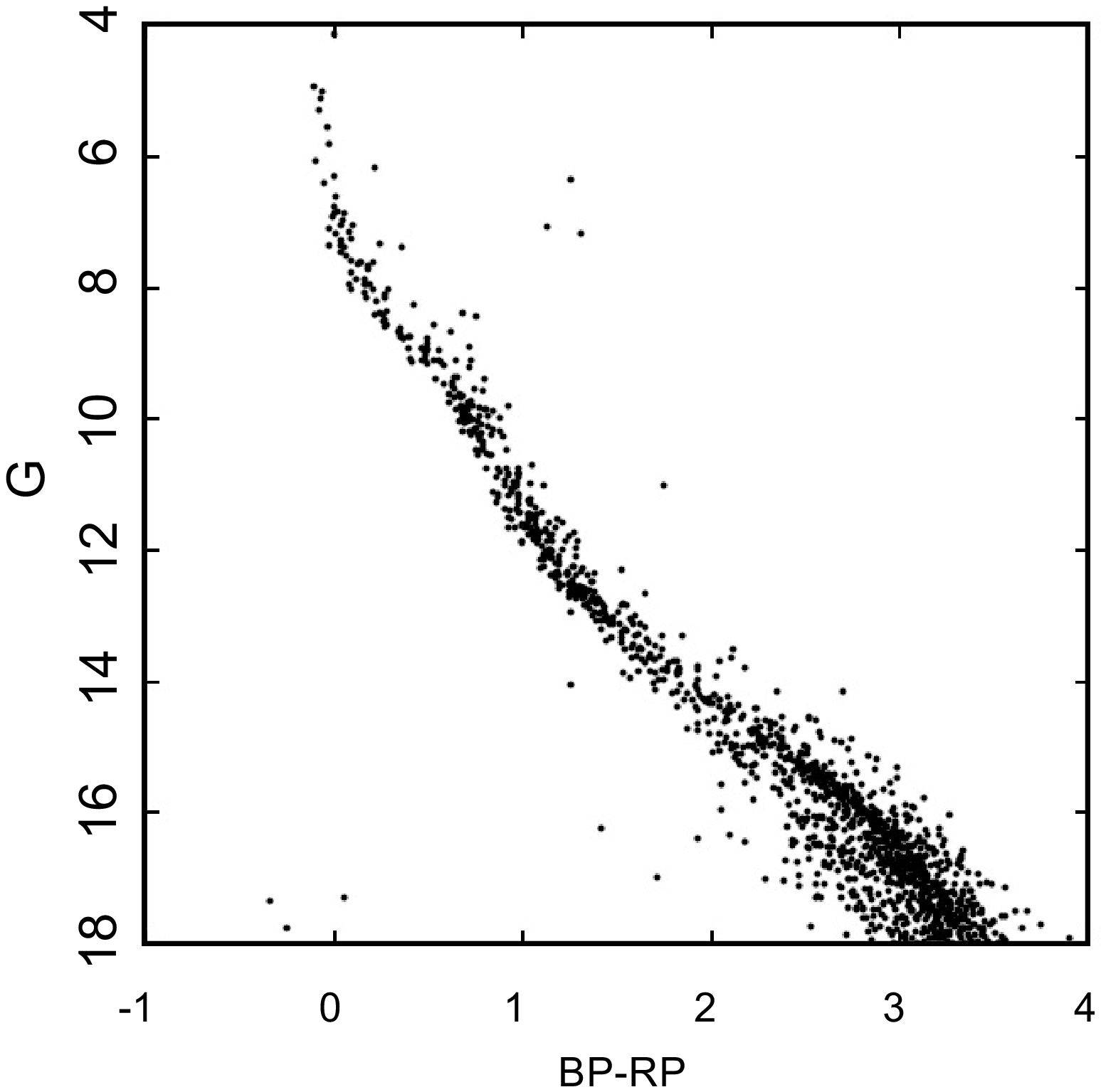}{0.4\textwidth}{(a)}
          \fig{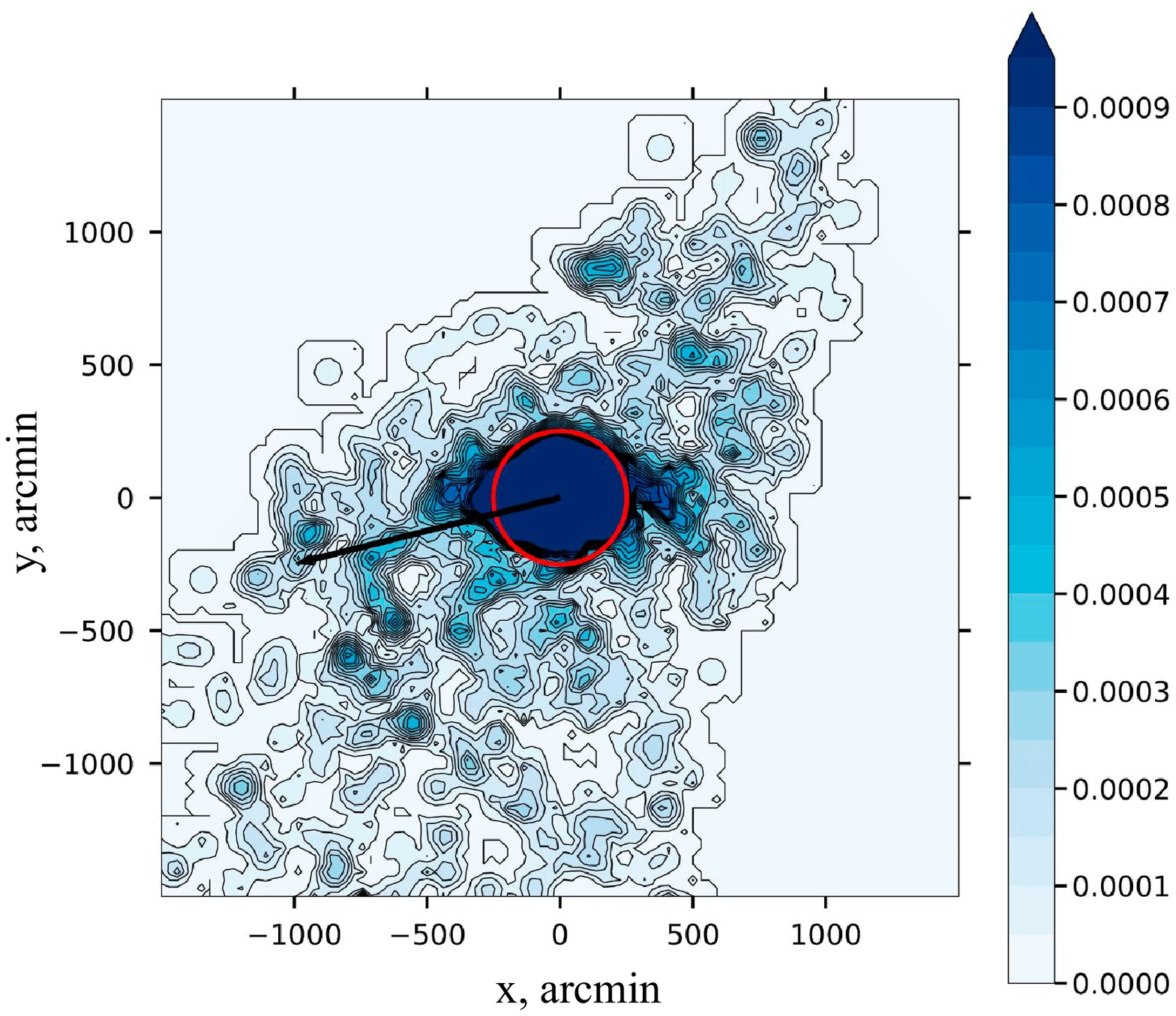}{0.45\textwidth}{(b)}}
\caption{Sample 2: (a) - Color-magnitude diagram; (b) - surface density map, designations are the same as in Fig.1.
The density values are in $\rm arcmin^{-2}$.}
\label{S2}
\end{figure}

Fig.1 shows the surface density map for Sample 1 obtained using a kernel density estimator (KDE) \citep{kernel}. This density map was plotted with the quartic kernel for two dimensions \citep{kernel} (just as all density maps in this paper) and the kernel halfwidth $h=120$ arcmin. In this map one can see two populations --- the cluster in the center and the stream (the structure elongated of about 70 degrees in approximately diagonal direction). The conclusion
that the stream could be a tidal feature associated with the cluster is not plausible since the stream direction is significantly different from the cluster motion.

Then, we made an attempt to get a cluster sample by gradually squeezing the parameters' limits
 $\pi$, $\mu_l$, and $\mu_b$ around their mean values ($\overline{\pi}=5.7$ mas, $\overline{\mu_l}=33$ mas/yr, $\overline{\mu_b}=-8.5$ mas/yr; these mean values were taken approximately from the diagrams `parameter-G'). We controlled this iterative process by monitoring visually the color-magnitude diagram (CMD) of the residual sample at every step. We terminated the process when the cluster sequence emerged clearly in the CMD against a minimum field star contamination. The final limits on the parallaxes and the proper motions are: $\pi\in[5.05;6.35]$ mas, $\mu_l\in[26;40]$ mas/yr, $\mu_b\in[-11;-6]$ mas/yr.

The left panel of Fig. 2 shows the CMD at the end of the process.
Some field stars are still present close to the cluster main sequence between G=15 mag and G=18 mag. This selection returned 1413 stars, and we will refer to it as Sample 2. Adopting the Padova suite of stellar models \citep{Padova}
and using cluster parameters as in Table 1, we
computed Sample 2 total mass as 924 $M_{\odot}$. We do not accompany this value with an uncertainty  since the
large part of it resides in the mass-luminosity relation used by \citet{Padova}, which remains unpublished.

For this mass, the tidal radius is estimated via the \citet{King} expression, where the Oort constants for
the Solar vicinity are taken from \citet{BoBa}. It amounts at $R_t=12.8\pm0.4$ pc. This is clearly a lower limit, because we applied a luminosity cut at
G$=$18 mag. The estimate of uncertainty is, also, a lower one, because we took into account only the uncertainties
on the Oort constants. The tidal-radius circle is indicated with a red  circle in Fig.1 and Fig.2 (right panel). The tangential velocity vector for the stars inside the tidal radius is shown as a black arrow.

Fig.2 (right panel) shows the surface density distribution for Sample 2 stars plotted by KDE with the quartic kernel halfwidth $h=100$ arcmin. We would like to highlight two evidences. First, the
stream stars did not disappear completely. This means that the cluster and the stream lie very close in 3-dimensional space of parallaxes
and proper motions, and one needs more sophisticated methods to separate
them. Second, we notice some hints of tidal features close to Alpha Persei tidal radius, which are not perfectly aligned with its velocity vector.

\section{Basic properties of the cluster and the stream}

In order to separate the cluster, the stream, and the field, we used DBSCAN \citep{DBSCAN}. \citet{Orion} and \citet{Vela} successfully used this procedure for selection of the filament-like structures in  regions of  active or recent star formation.
We run DBSCAN for three different configurations of the parameter space. The first one (C1) is a
five-dimensional ({\it l,b,$\pi$,$\mu_{\alpha}$, $\mu_{\delta}$}) space. The second one (C2) is a  three-dimensional ({\it $\pi$,$\mu_l$,$\mu_b$}) space, where $\mu_l$ and $\mu_b$ are proper motions
in Galactic coordinates. We refer to the third configuration (C3)
as the reduced three-dimensional volume ({\it $\pi$,$\mu_{lr}$,$\mu_{br}$}), where $\mu_{lr}$ and $\mu_{br}$ are
the residual proper motions with respect to the mean projected motion of
cluster stars (see above). Table 2 lists  the adopted DBSCAN  parameters and the results of the selection of stars in the cluster, in the stream, and
in the field (being the field the left-over after cluster and stream removal). $\varepsilon$ is the selection radius and $minPts$ is the minimum points number.

We explored DBSCAN parameters in a very wide limits' range ($\varepsilon\in[0.001;0.8]$, $minPts\in[10;1000]$). However, the separation of the cluster and the stream is achieved in a very narrow interval of  $\varepsilon$ and $minPts$ only, close to the values listed in Table 2. Even a small shift tended to situation when the cluster and the stream were not separated. The values of DBSCAN parameters were found by going over the $\varepsilon$ with some step and looking for an appropriate $minPts$.

\begin{table}
\normalsize
\bigskip
\begin{center}
\vspace{2 mm} Table 2. Parameters' values and results for the selection
with DBSCAN.

\vspace{2 mm}
\begin{tabular}{|l|c|c|c|}
\hline
Space         & $(l,b,Plx,\mu_\alpha,\mu_\delta)$ & $(Plx,\mu_l,\mu_b)$ & $(Plx,\mu_{lr},\mu_{br})$ \\
\hline
Configuration   &   C1  &   C2  &   C3  \\
\hline
$\varepsilon$   & 0.2   &  0.05 &  0.05 \\
$min(Pts)$      & 400   &  450  &  600  \\
Cluster (stars) & 2147  & 1884  & 1370  \\
Stream (stars)  & 3881  & 9197  & 2977  \\
Field (stars)   & 26414 & 21361 & 28095 \\
\hline
\end{tabular}
\end{center}
\end{table}

\begin{figure}
   \centering
   \includegraphics[width=16truecm]{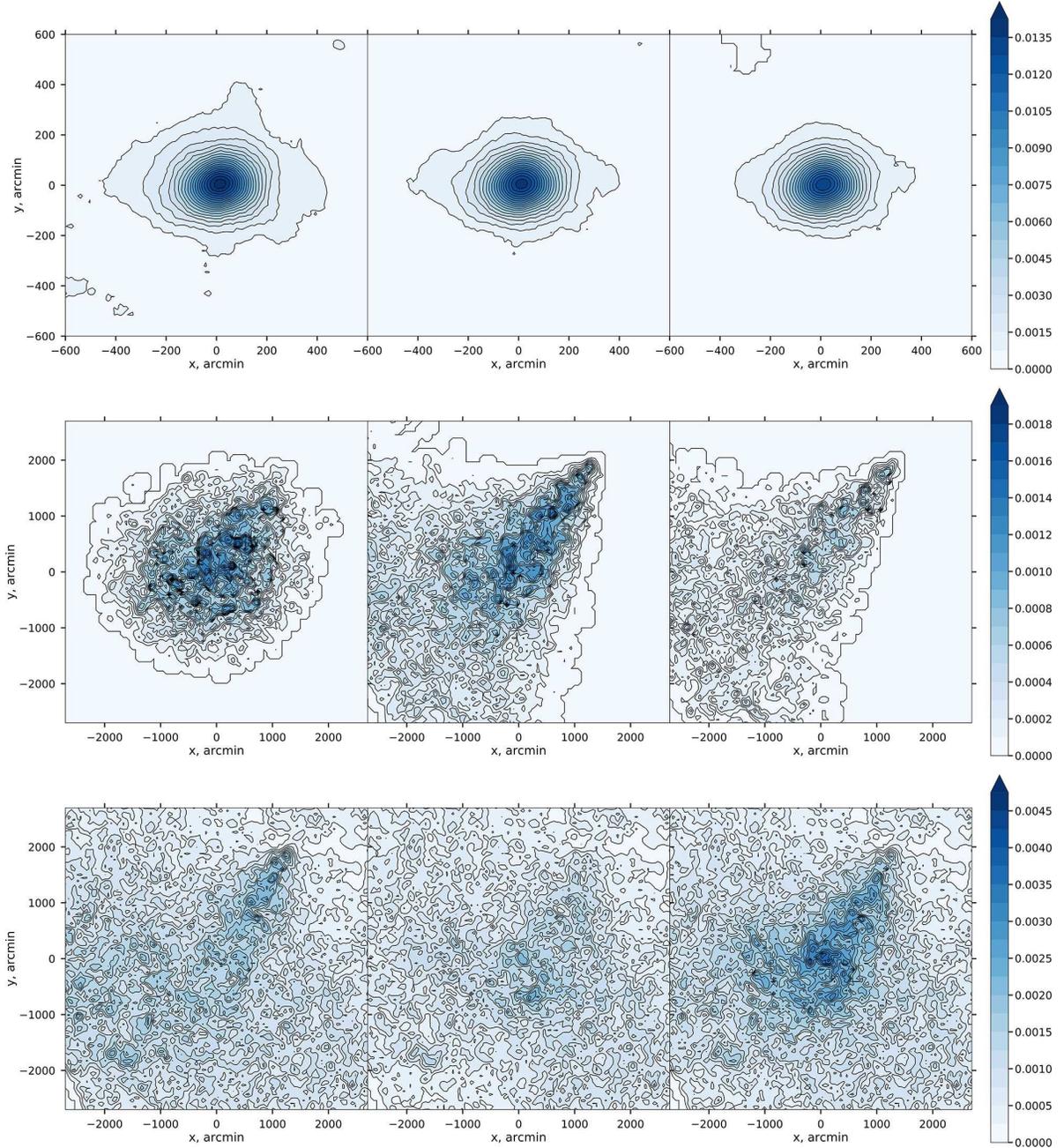}
   \caption{Surface density maps for the cluster (upper panels), the stream (central panels), and the field (lower panels).  From the left to the right panels refer to C1, C2, and C3 DBSCAN configurations. Density is in $\rm arcmin^{-2}$.}
   \label{DBSCAN_res}
   \end{figure}

\begin{figure}
   \centering
   \includegraphics[width=16truecm]{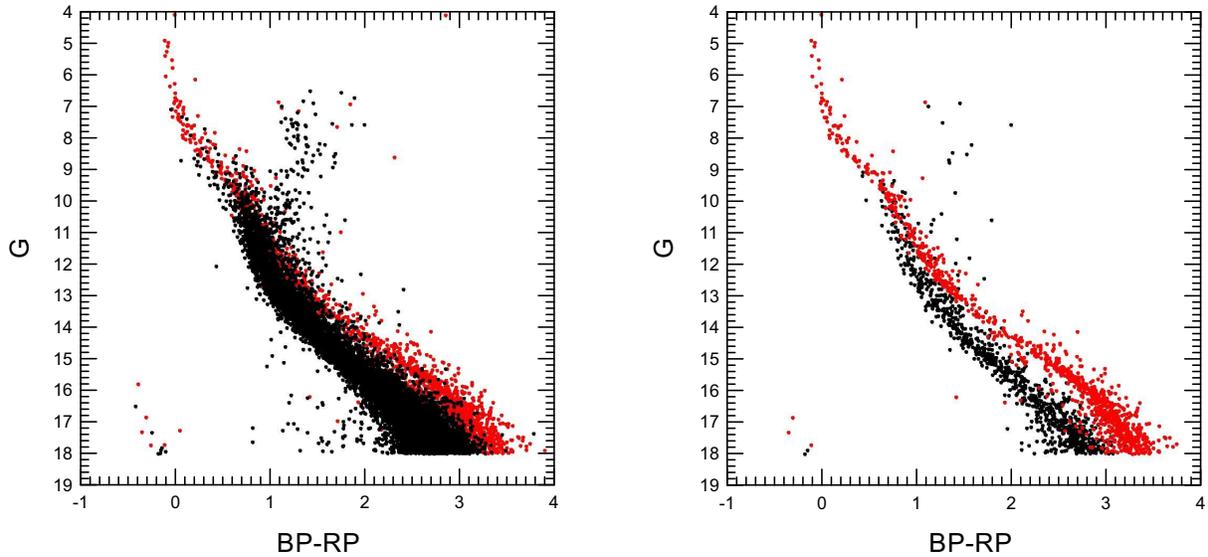}
   \caption{The color-magnitude diagrams (CMDs) for the cluster (red points) and for the stream (black points). The left panel shows the stars selected in C2 parameter space. The right panel shows the CMD for stars in common in all the 3 different configurations.}
   \label{CMDs}
   \end{figure}

\begin{figure}
   \centering
   \includegraphics[width=7truecm]{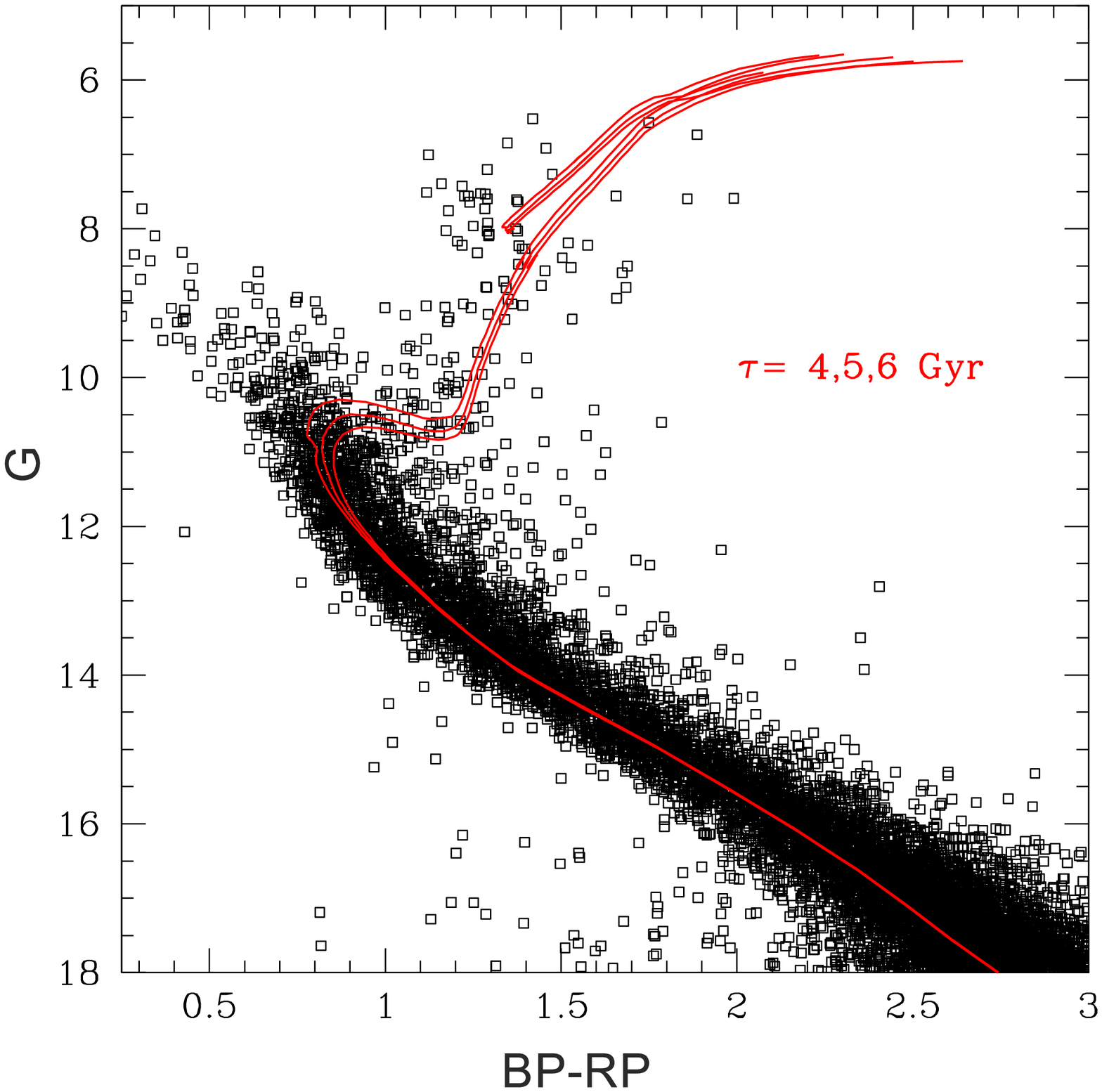}
   \caption{The color-magnitude diagram (CMD) for the stream (C2 parameter space). The red lines show the isochrones for solar metallicity and the age of 4,5,6 Gyrs.}
   \label{CMD+iso}
   \end{figure}

The outcome of the three different selection criteria (C1, C2, and C3) can be analyzed using stars' surface density maps (see Fig.~3). All these maps were plotted by KDE with the kernel halfwidth of $h=120$ arcmin. The effectiveness of the selection in particular can be tested against the general field
stars' distribution, which, ideally, should be uniform.
The selection C2 clearly provides the best result from this point of view. The cluster corona is elongated along the Galactic longitude. However, in the inner part of the cluster, the density contours are aligned with the cluster velocity vector (see Fig.1 for a comparison).

In Fig.4 (left panel) we show the  CMDs for the cluster (color coded in red) and the stream (color coded in black)
for the best performing configuration C2. The right panel, on the other hand, shows the CMD for the stars in common in all the three different DBSCAN configurations.

We then focused on the stream itself, and attempt at determining its fundamental parameters as a stellar population.
In Fig.~5 we compare the star distribution in the CMD with isochrones from the Padova suite of models\citep{Padova}.
Given the proximity of this structure to the Sun, we explored solar metallicity models. The fitting has been performed eyeballing the isochrone with respect to the star distribution in the vicinity of the turn off point and in the red clump region. We found a plausible age of $5\pm1$ Gyr.

To obtain this age estimate, we shifted the isochrones horizontally by  $E(B-V) \sim0.1$ (see Table 1), which gives $E(BP-RP)=0.13\pm0.03$ according to \citet{Cardelli+1989} and \citet{O'Donnell1994} extinction curve. This value of the reddening $E(B-V)$ is confirmed by the analysis of the 2MASS \citep{2MASS} color-color diagram J-K vs H-K. The reddening for the candidate stream stars (in accordance with the C2 DBSCAN configuration) as derived from the aforementioned diagram turns out to be $E(B-V)=0.09\pm0.06$ mag.
The vertical shift yields for the stream a distance modulus $(m-M)=7.4\pm0.2$ mag and $(m-M)_0=7.1\pm0.2$ mag with $A_G=0.27$ mag (corresponding to the same extinction curve). It corresponds to a mean heliocentric distance of  263$^{+25}_{-23}$ pc.

Then, from the inspection of these CMDs, we can conclude that the stream consists of a stellar population $\sim5$ Gyr old, significantly older than the cluster. This rules out a common origin scenario for the cluster and the stream. The most plausible explanation for the stream could be that
it is the leftover of an old disrupted star cluster moving in the vicinity of the Alpha Persei cluster.

Besides, we also note that the stream appears to lie generally behind the cluster. The difference between the adopted heliocentric distance of the cluster and the mean heliocentric distance of the stream is $\approx$90 pc.
Finally, by looking more closely at the left panel of Fig.4 one can notice from the MS width that the stream looks extended along the line of sight by about 180 pc, from 190 to  370 pc from the Sun, approximately. These estimates have been obtained from the main sequences band boundaries at e.g. (BP-RP)=2 mag, assuming the same extinction for the cluster and for the stream. A more detailed  investigation of the spatial structure of the stream shows that this structure is more complicated.

\begin{figure}
   \centering
   \includegraphics[width=9truecm]{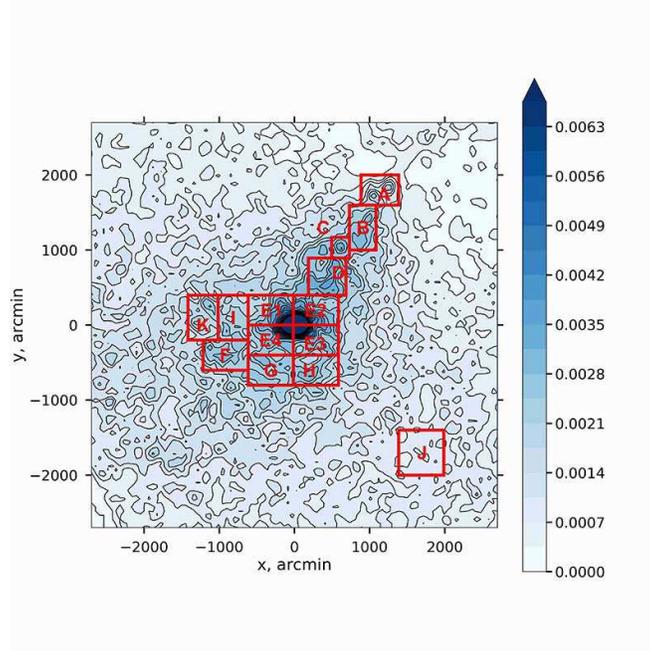}
   \caption{Selected regions to analyze the structures along the line of sight.}
   \label{regions}
   \end{figure}

\begin{figure}
   \centering
   \includegraphics[width=12truecm]{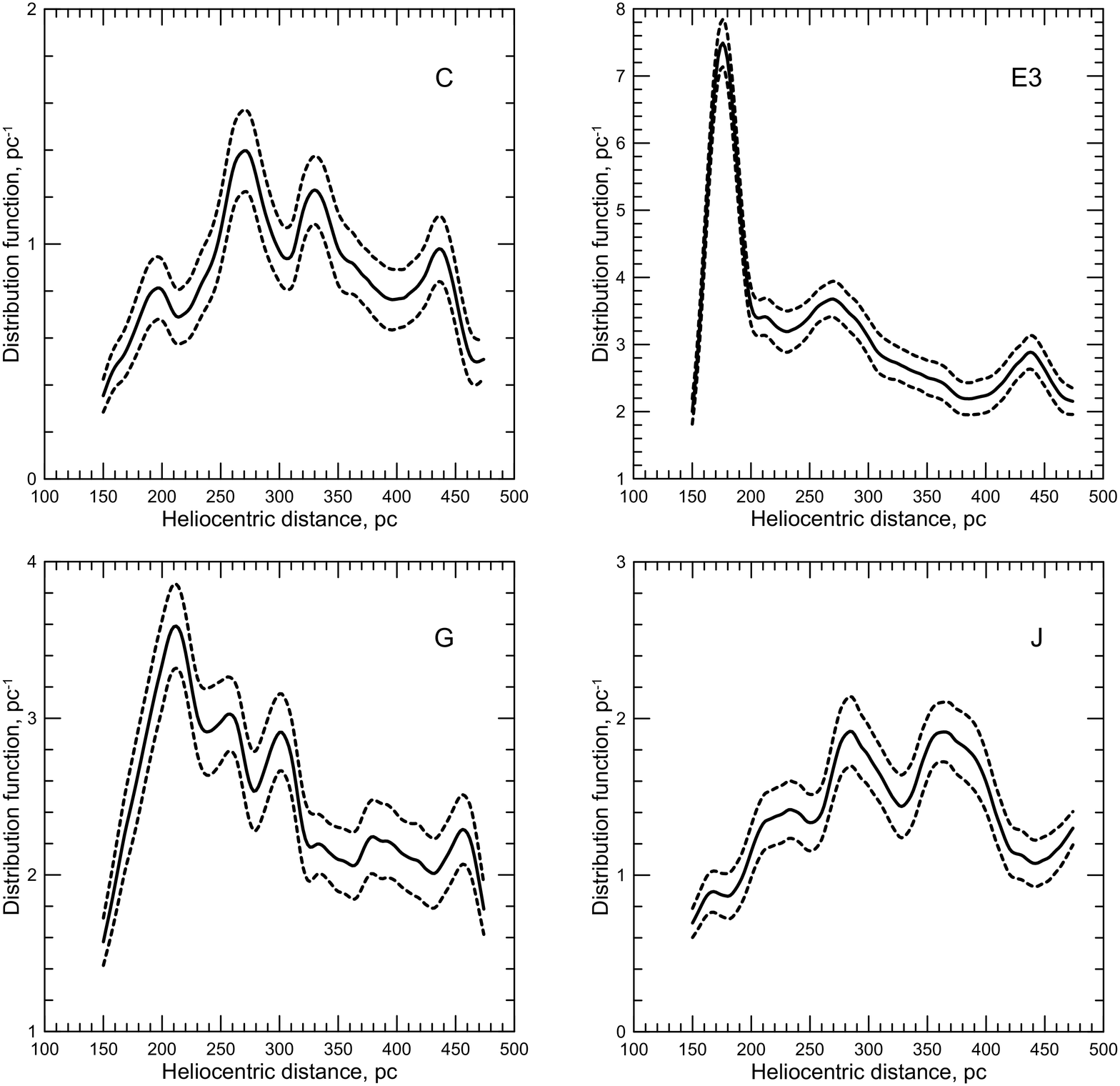}
   \caption{The distance distributions for regions C, E3, G,and J. The solid lines show the distance distributions, and the dotted lines show the 2-$\sigma$ confidence interval plotted by a `smoothed bootstrap' method \citep{MT,kernel}. }
   \label{distances}
   \end{figure}

The existence of the stream as a spatially confined structure is confirmed by the distribution of stellar heliocentric distances for a series of regions across the stream area that we outlined by iso-density contours (see Fig.\ref{regions}). Fig.\ref{distances} shows such distributions plotted with KDE  (quartic kernel for one dimension with the kernel halfwidth of $h=25$ pc) for regions C, E3 (this region includes the cluster), G, and J (this region represents the general field).

In order to derive these distributions we downloaded from Gaia DR2 a new sample (Sample 3), which is the same as Sample 1 except for the parallax limit $Plx\in[2;8]$ mas, and therefore includes more distant stars. This Sample 3 contains 59803 stars having $G<18$. When analyzing the distance distributions one needs to remember that within the same solid angle and with the uniform spatial distribution of stars the number of stars will increase with the distance. So, the increase of the number of stars at increasing distance for region J is due to this geometric effect.

We calculated the distances by inverting parallaxes. We are aware that the direct inversion of the parallax is only a crude approximation of the distance \citep{Luri}, but we expect that for the small distance range we are working with the implied error are not very important. Parallaxes in Sample 3 have indeed small relative errors: 90 percent of stars have the relative parallax error less than 0.06 and 97.7 percent of stars have this error less than 0.1.

These distance distributions show several interesting features. Firstly, the stream is standing against the background as a net overdensity. Secondly, the lower-left part of the stream is closer to us than the upper-right part of the stream (the density maximum in region G is at 210 pc and in region C at 270 pc; the density maximum for the cluster in region E3 is at 176 pc which is in an excellent agreement with the cluster data). Thirdly, the stream is clumpy. It is well seen on the density maps (Fig.\ref{map_S1} and Fig.\ref{DBSCAN_res}) and on distance distributions which often have  multiple maxima. Fourthly, the stream is partially overlapping with the cluster. The width of the stream along the line of sight is in good agreement with the one inferred from the CMD (see above).

\begin{figure}
   \centering
   \includegraphics[width=17truecm]{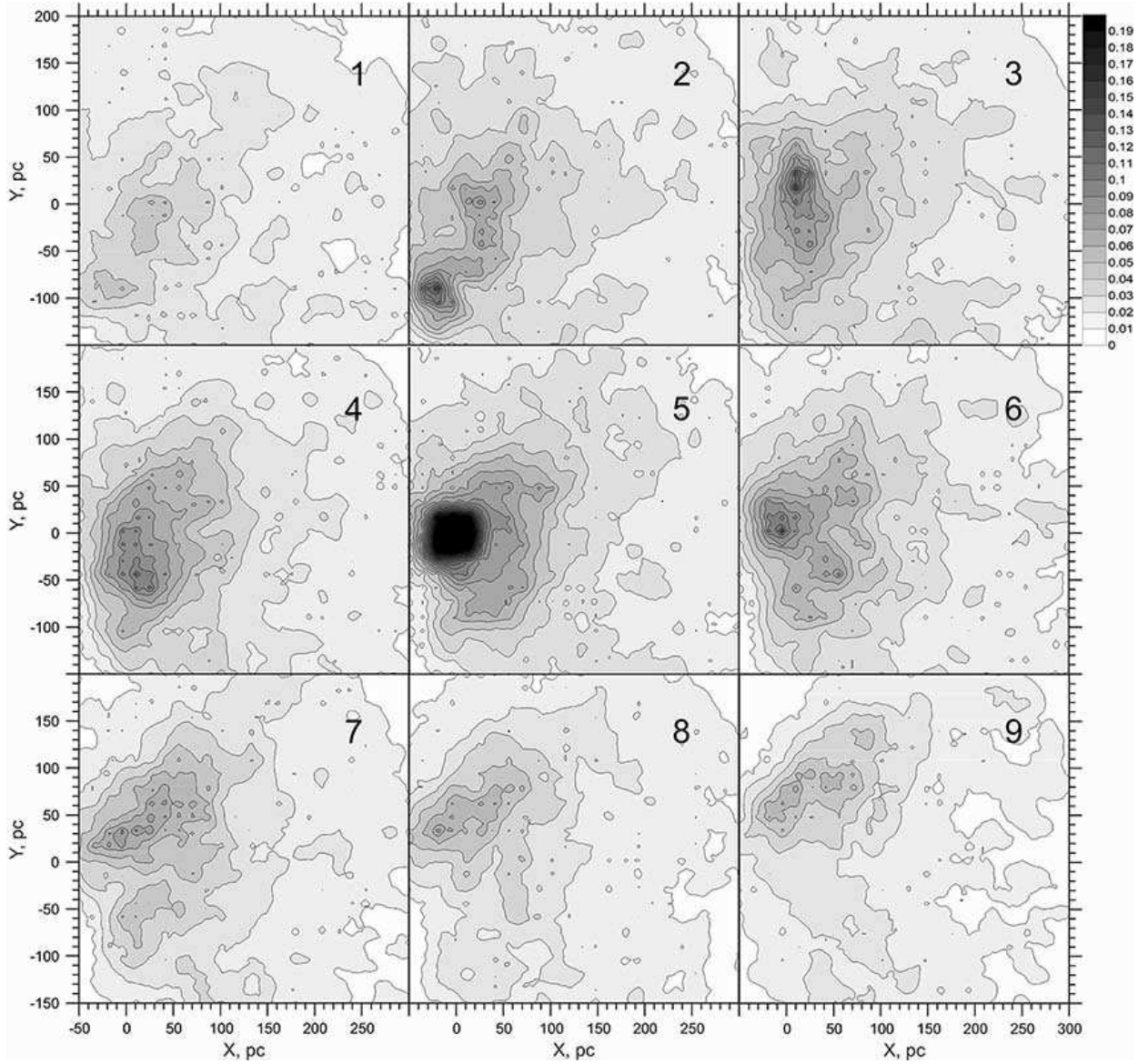}
   \caption{The density distributions in the projection onto XY plane for 20-pc-thick slices along Z axis. The density units are pc$^{-2}$. 1 --- $Z\in[-90;-70]$ pc, 2 --- $Z\in[-70;-50]$ pc, 3 --- $Z\in[-50;-30]$ pc, 4 --- $Z\in[-30;-10]$ pc, 5 --- $Z\in[-10;10]$ pc, 6 --- $Z\in[10;30]$ pc, 7 --- $Z\in[30;50]$ pc, 8 --- $Z\in[50;70]$ pc, 9 --- $Z\in[70;90]$ pc. }
   \label{slices}
   \end{figure}

With the goal to provide a sharper view of  the structure of the stream we calculated coordinates X, Y, and Z for stars of Sample 3.
The center of this coordinate system coincides with the cluster center, while the XY-plane is parallel to Galactic plane. The X axis is directed towards the Galactic anti-center, the Y axis follow the tangent of the cluster Galactic circular orbit, and the Z axis is perpendicular to Galactic plane. Fig.\ref{slices} shows the surface density distributions in a projection onto the XY-plane for 20-pc-thick `slices' along Z axis (KDE with the quartic kernel halfwidth of $h=20$ pc). The density levels in units of pc$^{-2}$ were selected to provide the best contrast and highlight the structure of the stream effectively. It is clear that the stream has a complicated structure, it consists of several clumps, and is overlapping with the cluster. Anyway, our previous conclusions on the overall geometry of the stream are confirmed.

\begin{figure}
   \centering
   \includegraphics[width=17truecm]{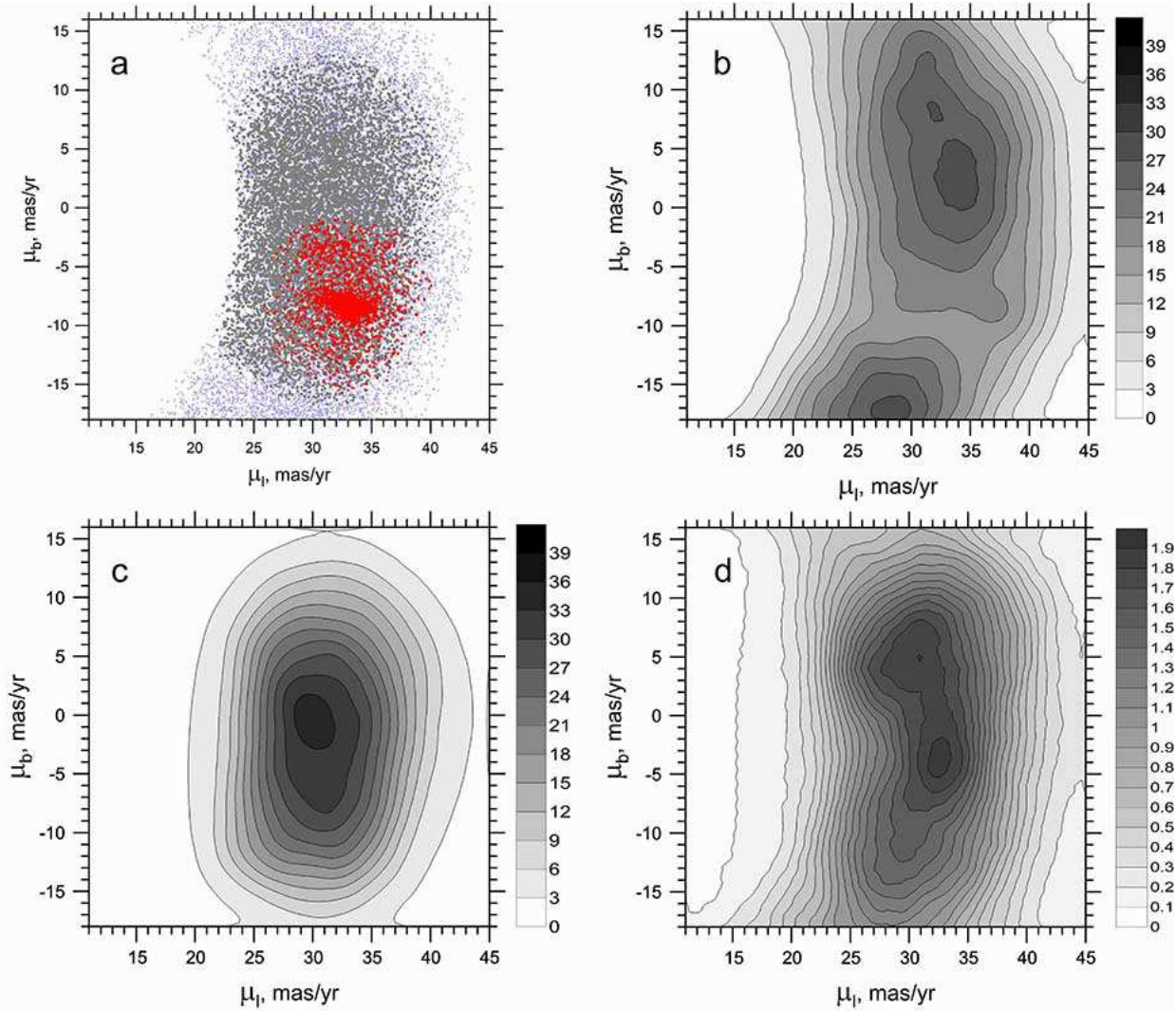}
   \caption{The proper motion diagrams for the selected subsystems. (a) The cluster (red dots), the stream (grey dots), and the field (violet dots) in accordance to C2 DBSCAN configuration. (b) -- (d) The density maps in the proper motion plane for the field (b), the stream (c), and the WD candidate sample (d) plotted by KDE with the quartic kernel halfwidth of $h=3$ mas/yr. }
   \label{PM}
   \end{figure}

\begin{figure}
   \centering
   \includegraphics[width=18truecm]{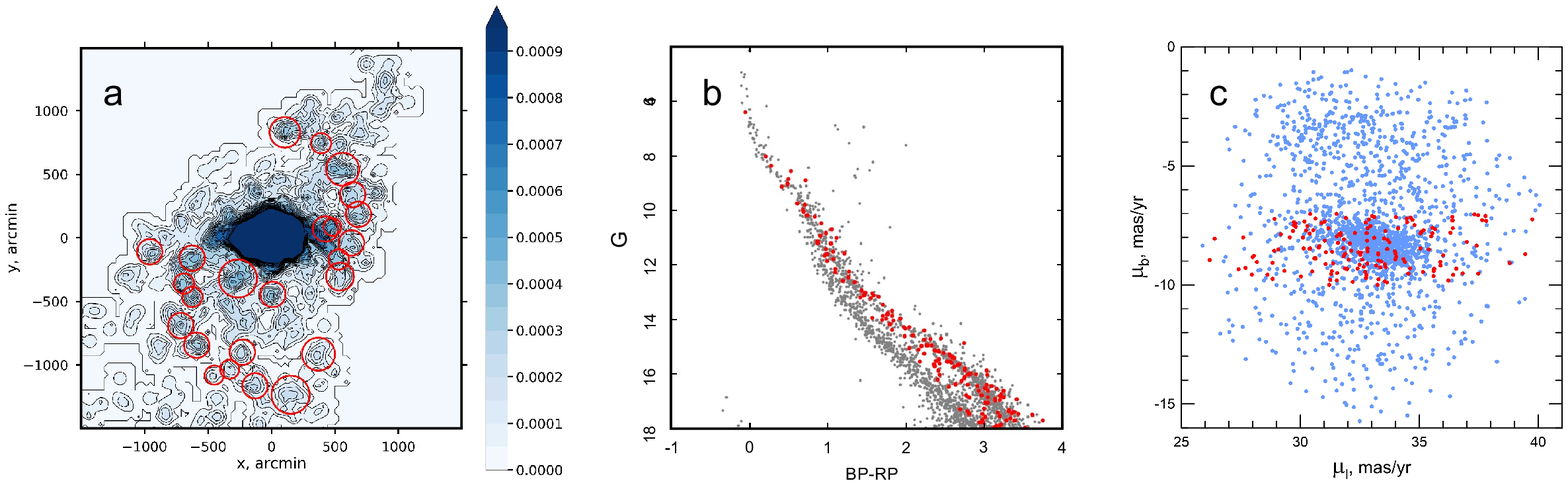}
   \caption{ (a) The density map of the Sample 2 with the red circles marking the subgroups of the ``tidal tails'' (designations are the same as in Fig.2b). (b) The CMD for these stars (red points) comparing with the CMD position of the cluster and the stream stars (see the right panel of Fig.4). (c) The proper motion diagram for the cluster (C2) stars (blue dots) and stars of the ``tidal tails'' (red dots).}
   \label{tidal_tails}
   \end{figure}

An upper limit of the stream mass is $\sim 6000 M_{\odot}$. This
has been obtained by merely counting the number of stream stars (C2) and it is an upper limits since we cannot exclude some amount of field star contamination. Therefore, the stream seems to identify the left over of quite a massive cluster which is now on the verge to dissolve into the Galactic field.

To lend further support to this hypothesis we checked the distribution of the various subsystems in the proper motion plane (vector point diagram). Fig.\ref{PM}a shows three subsystems selected with the {\it C2} DBSCAN configuration:   violet dots denote the field,  gray dots  the stream, and red dots the cluster. In order to make the difference between the field and the stream more evident we plotted the density maps in a vector point diagram for these subsystems ( Fig.\ref{PM}b and  Fig.\ref{PM}c, respectively). It is readily seen that the stream has a well-marked central concentration unlike the field. The dispersion of proper motions for the stream is larger than for the cluster. This picture resembles a disrupting star cluster. The kernel half-width for these maps was taken to be $h=3$ mas/yr.

Fig.\ref{PM}d shows the distribution of the white dwarf (WD) candidates (see below) in the proper motion plane. This distribution shows two maxima. The upper one is close to the upper maximum of the field distribution (Fig.\ref{PM}b), and the lower one is close to the maximum of the stream distribution (Fig.\ref{PM}c). We stress here that the WD sample consists of stars having $G>18$ mag (Sample 1a, see below). These stars have usually large errors of proper motions. Consequently, it is difficult to make the firm conclusions only considering the proper motion distribution of these stars.

We could not find a set of DBSCAN parameters suitable to isolate  the {\it tidal tails} structure we have seen in Fig.\ref{S2} as belonging to the cluster. In order to clarify the nature of this structure, we compiled a sample of stars (called hereafter as TTS) in a following way. We took stars of Sample 2 which formed the {\it tidal tails} structure in the projection to a tangent plane, namely, the stars inside red circles at Fig.\ref{tidal_tails}a. Nearly all stars of the TTS sample lie outside the cluster sample being selected by the DBSCAN procedure.

Fig.\ref{tidal_tails}b shows the CMD, where the grey points are from the right panel of Fig.4 (the cluster and the stream stars in common for all the three DBSCAN selection criteria) and the red points correspond to stars which lie inside the encircled regions in Fig.\ref{tidal_tails}a (TTS). The vast majority of the red points lie very close to the cluster sequence, with a few exception of stars scattered away from the MS. This fact supports the origin of {\it tidal tails} stars from the cluster.

An additional support to a hypothesis that the TTS stars originated from the cluster is in the proper motion distribution. Fig.\ref{tidal_tails}c shows the proper motion diagram for the cluster stars (C2 DBSCAN configuration, blue dots) and stars of the TTS sample (red dots). The stars of the TTS sample lie well inside the region occupied by the cluster stars. Thereby, we can confidently claim that the tidal-tail-like structures from Fig.\ref{S2} are real and most probably connected with the cluster.

A possible explanation why the DBSCAN did not add the TTS stars to the cluster sample is that in fact these structures are poorly-populated with respect to other features.

What is the origin of these {\it tidal tails}? A gravitational interaction with the stream seems unlikely due to low mass of the stream. \citet{Danilov1994,Danilov&Seleznev1995a,Danilov&Seleznev1995b} showed that massive gas-stellar complex (with the mass of $10^5-10^8$ solar masses) can exert a prominent gravitational influence on an open star cluster. Gould Belt with the mass of about $10^6$ solar masses could play such role, but Alpha Persei open cluster is very close to its center (see a review in \citet{GouldBelt}). Our mind is that these {\it tidal tails}, although not perfectly aligned with Alpha Persei velocity vector, are most probably of Galactic origin, in analogy with other cases reported in the literature \citep{Rup147}.

In Table 3 we list the mean velocities of the cluster stars, the mean residual velocities of the stream stars, and the mean velocities of the {\it tidal tails} stars (residual velocity with respect to the cluster). The mean velocity of the {\it tidal tails} are very close to the mean velocity of the cluster stars. The dispersion, though, is high.

\begin{table}
\normalsize
\bigskip
\begin{center}
\vspace{2 mm} Table 3. Mean velocities of the different populations in the vicinity of Alpha Persei cluster.

\vspace{2 mm}
\begin{tabular}{|l|c|c|c|c|c|}
\hline
Population  &  Parameter   & \multicolumn{4}{c|}{Selection with DBSCAN}                 \\
\hline
            &                &      C1       &       C2      &      C3       & Intersection \\
\hline
            &$\mu_l$, mas/yr & 32.7$\pm$0.1 & 32.9$\pm$0.1 & 33.0$\pm$0.1 & 33.1$\pm$0.4 \\
\cline{2-6}
Cluster     &$\mu_b$, mas/yr & -7.3$\pm$0.1 & -7.7$\pm$0.1 & -7.8$\pm$0.1 & -7.9$\pm$0.1 \\
\cline{2-6}
            &  $V$, km/s     & 29.2$\pm$0.1 & 28.4$\pm$0.1 & 28.3$\pm$0.1 & 28.4$\pm$0.0 \\
\hline
            &$\mu_l$, mas/yr & 26.5$\pm$0.1 & 30.8$\pm$0.0 & 30.5$\pm$0.1 & 28.1$\pm$0.1 \\
\cline{2-6}
Stream      &$\mu_b$, mas/yr & -5.7$\pm$0.1 & -2.0$\pm$0.1 & -0.9$\pm$0.1 & -2.9$\pm$0.1 \\
\cline{2-6}
            &    $V$, km/s   & 35.3$\pm$0.1 & 42.3$\pm$0.1 & 44.4$\pm$0.1 & 40.7$\pm$0.1 \\
\cline{2-6}
            &   $V_r$, km/s  &  6.9$\pm$0.1 & 13.9$\pm$0.1 & 16.0$\pm$0.1 & 12.3$\pm$0.1 \\
\hline
Tidal tails &    $V$, km/s   &              &              &              & 28.3$\pm$3.2 \\
\cline{2-6} &   $V_r$, km/s  &              &              &              & 0.04$\pm$3.15\\
\hline
\end{tabular}
\end{center}
\end{table}

Table 3 also illustrates that all three different DBSCAN selections yield similar results concerning the cluster but different results concerning the stream (this is clearly seen also in Fig.3).

\section{White dwarfs of the stream}
To lend further support to our findings we looked at the space distribution of white dwarf (WD) stars in the area.
In fact, the stream is old and populous enough to harbour a noticeable population of WDs. Since we do not expect WDs from Alpha Persei, these stars should be genuine tracers of the stream.
Their distribution should therefore confirm both the structure and the age of the stream.

\begin{figure}
   \centering
   \includegraphics[width=16truecm]{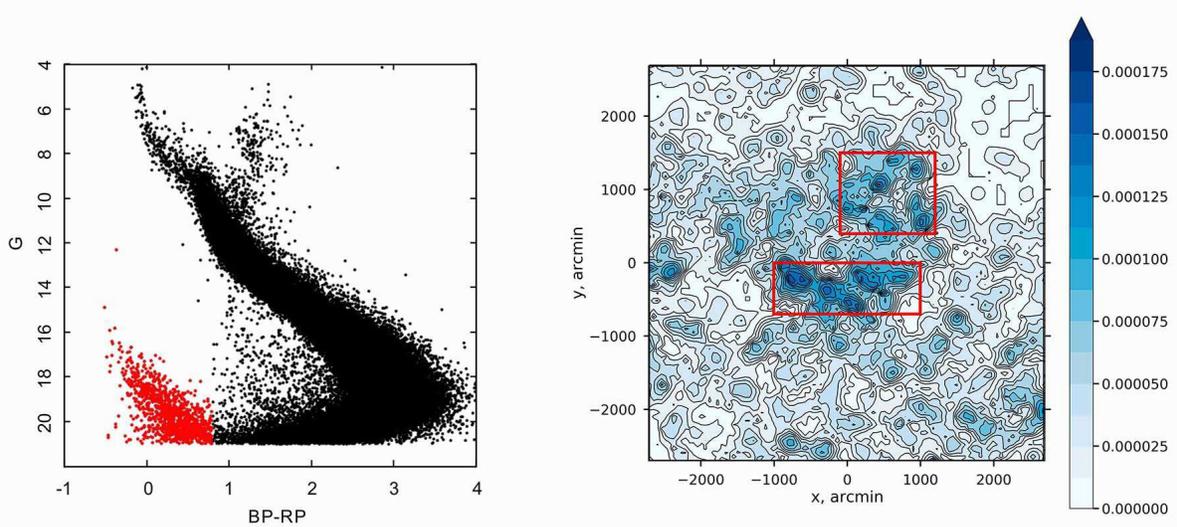}
   \caption{The left panel shows the CMD of the Sample 1 without the limit on the stellar magnitude (Sample 1a) with WD candidates marked by red. The right panel shows the density map for WD candidates. For stars from the red rectangles the distance distributions are shown in Fig.\ref{WDdist}. }
   \label{wd}
   \end{figure}

Unfortunately, Sample 1 contains very few WDs. Most WDs have in fact $G>18$ mag. Because of this,  to isolate the WD population of the stream, we turned to Sample 1a relaxing the magnitude constraint. Recall that this sample contains 60603 stars with the same limits on parallaxes and proper motions as Sample 1, but with G magnitudes up to the limit of the Gaia DR2 catalog. The left panel of Fig.\ref{wd} shows the CMD of this sample with WD candidates highlighted in red.
We also compared the expected position of WDs with the work of \citet{GaiaDR2_CMD}.
We relaxed the limit on the stellar magnitude because WD candidates with $G\leq 18$ mag were very few and, besides, did not show any spatial concentration.

\begin{figure}
   \centering
   \includegraphics[width=14truecm]{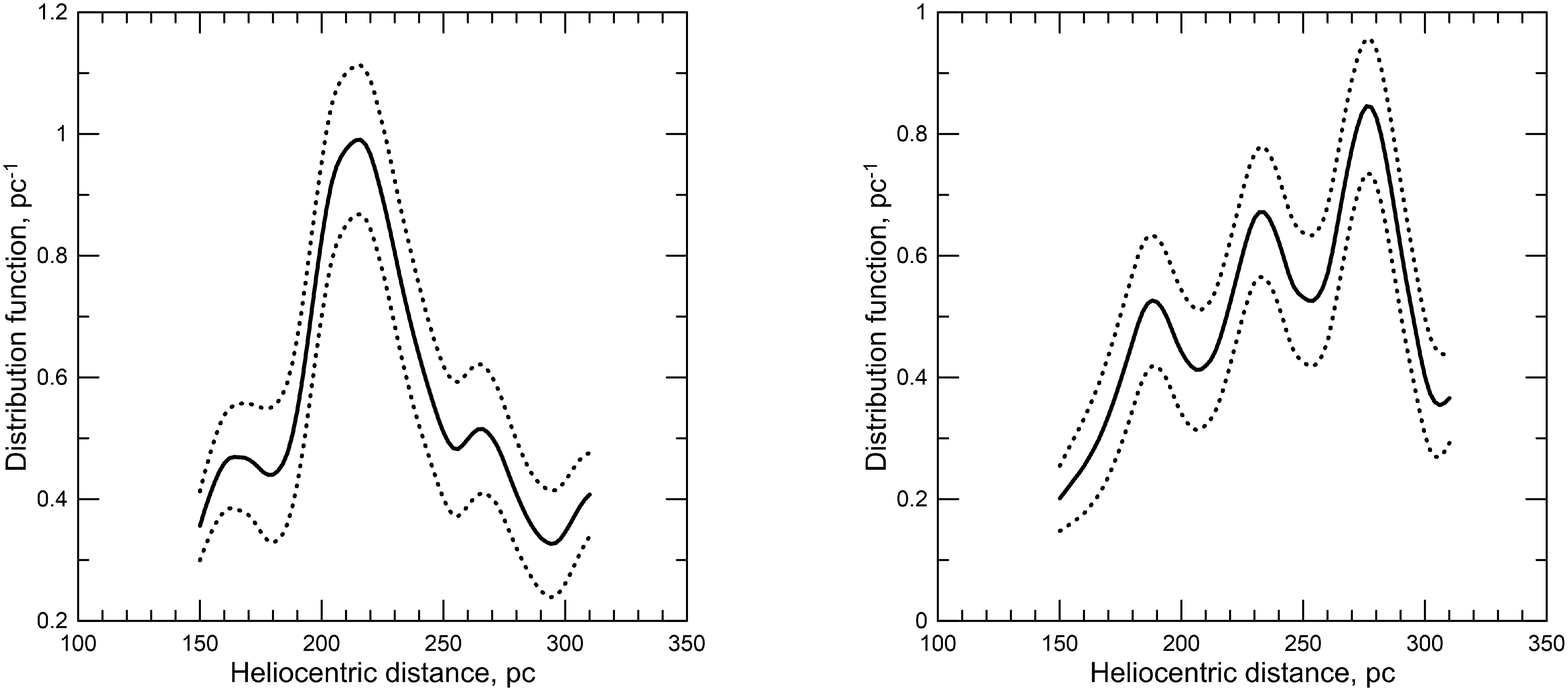}
   \caption{The left panel shows the distance distribution for the WD candidates from the lower red rectangle of Fig.\ref{wd}, right. The right panel shows the distance distribution for the WD candidates from the upper red rectangle of Fig.\ref{wd}, right. The designations are the same as in Fig.\ref{distances}.} 
   \label{WDdist}
   \end{figure}

The right panel of Fig.\ref{wd} shows the density map for the
CMD-selected WDs (KDE with the quartic kernel halfwidth of $h=230$ arcmin). The concentration of stars close to the stream region is readily seen. The depression (low density region) at the cluster position is due to the screening effect of the cluster which lies in front of the stream along the light of sight.

In order to confirm the spatial association of WDs to the stream, we selected two regions with a maximum surface density of the WD stars (the red rectangles in the right panel of Fig.\ref{wd}) and plotted the distance distributions of stars from these regions. We present these distributions in Fig.\ref{WDdist}. These distributions were plotted in the same way as Fig.\ref{distances}. The distance distributions of WDs show  distinct maxima. We can compare these distributions with Fig.\ref{distances}. The maximum in the left panel of Fig.\ref{WDdist} (for a lower rectangle of Fig.\ref{wd}, right panel) coincides well with the maximum for field G of Fig.\ref{distances}. The maxima in the right panel of Fig.\ref{WDdist} (for an upper rectangle of Fig.\ref{wd}, right panel) coincides well with the maxima for field C of Fig.\ref{distances}. We can note that the lower rectangle corresponds roughly to regions F,G,H of Fig.\ref{regions}, and the upper rectangle corresponds roughly to region C of Fig.\ref{regions}. Thus, the distance distributions of WDs correspond well to the distance distributions of the stream stars.

Finally, in Fig.\ref{PM}d we showed the density distribution of the WDs (selected as described above) in the proper motion plane. Unfortunately, the errors of the proper motions for stars with $G>18$ mag are large, and it is difficult to make any firm conclusions about a similarity of distributions in Fig.\ref{PM}b,c,d. Nevertheless, the visual comparison of these figures argue that the WD proper motion distribution shares the features of both the field and the stream (see discussion above in previous section).

Anyway, the spatial concentration is the clear evidence that most of the selected WDs are members of the stream. Also, it is a confirmation of the age estimate of the stream of  $\sim$5 Gyr.

\section{Conclusions}

In this work we investigated in details the projected  spatial distribution of stars in the surroundings of the Alpha Persei star cluster with the aim of isolating the stellar stream with which the cluster is mixed. The evidence and existence of
this stream have been repeatedly discussed in recent literature.

We succeeded to separate the cluster stars and the stream stars from the general Galactic field using DBSCAN. The stream exhibits a large radial extent (about 180 pc). The distance between the cluster center and the stream central line is about 90 pc; the stream lies generally in the background of the cluster and has a clumpy structure. The stream is significantly older than the cluster with an age of $\sim$5 Gyr. The presence of a conspicuous WD population that follows the stream spatial distribution lends further support to the age estimate of this structure. We estimated the stream mass to be $\sim 6000 M_{\odot}$.

The most probable interpretation of our results is that the stream is the relict of an initially massive star cluster that the Galactic tidal forces are bringing to the brink of dissolution. This point of view does not contradict the distribution of the proper motions of the stream stars.

The density map of Sample 2 (taken with very narrow limits on parallaxes) shows a presence of structures resembling tidal tails. An analysis of the proper motions and photometric properties of stars composing these tidal tails confirmed a very probable origin of these stars from the cluster. It is conceivable to envisage that the stream exerted a significant tidal action on the cluster, if it would be much more massive \citep{Danilov1994,Danilov&Seleznev1995a,Danilov&Seleznev1995b}. However, by considering similar cases in the solar vicinity, we favour a scenario in which Alpha Persei tidal tails are of Galactic origin.

\acknowledgments
This work was supported by the Ministry of Science and Higher Education of the Russian Federation, FEUZ-2020-0030, and by the Act no. 211 of the Government of the Russian Federation, agreement no. 02.A03.21.0006.
This work has made use of data from the European Space Agency (ESA) mission
{\it Gaia} (\url{https://www.cosmos.esa.int/gaia}), processed by the {\it Gaia}
Data Processing and Analysis Consortium (DPAC,
\url{https://www.cosmos.esa.int/web/gaia/dpac/consortium}). Funding for the DPAC
has been provided by national institutions, in particular the institutions
participating in the {\it Gaia} Multilateral Agreement. The input of the anonymous referee has been greatly appreciated.

{}

\end{document}